\documentclass[a4paper,12pt]{article}
\pdfoutput=1
\usepackage{cite}
\usepackage{jheppub}
\usepackage{hyperref}
\hypersetup{colorlinks=true}
\usepackage[T1]{fontenc} 
\title{Neumann-Rosochatius system for strings on I-brane }
\author{Adrita Chakraborty, Nibedita Padhi, Priyadarshini Pandit, Kamal L. Panigrahi}
	 \affiliation{Department of Physics, Indian Institute of Technology Kharagpur-721302, India}
	\emailAdd{adimanta09@iitkgp.ac.in}
	 \emailAdd{nibedita.phy@iitkgp.ac.in}
	 \emailAdd{pandit006@iitkgp.ac.in}
	\emailAdd{ panigrahi@phy.iitkgp.ac.in}
\abstract{We study rigidly rotating and pulsating strings in the background of a 1+1 dimensional intersection of two orthogonal stacks of fivebranes in type IIB string theory by using the Neumann-Rosochatius (NR) model. Starting with the Polyakov action of the probe fundamental string we show that a generalised ansatz reduce the system into the one dimensional NR model in the presence of flux. The integrable construction of the model is exploited to analyze the rotating and oscillating string solution. We render the large $J$ BMN-type expansion for the energy of rotating string states while the same for the oscillating string has been derived in long string and small angular momenta limit.}


\begin{document}
\maketitle
	\flushbottom
	\section{Introduction} 
	Appearance of classical integrability in the famous AdS/CFT correspondence \cite{Maldacena:1997re,Gubser:1998bc, Witten:1998qj} 
	has paved a path to cope with the long-standing quest of matching the energy of free string states with anomalous dimensions of different CFT operators for any arbitrary 't Hooft coupling. As any integrable physical system is believed to acquire infinite tower of conserved quantities in involution, it provides an elegant mathematical technique for exactly solving the corresponding system. Integrability was first developed in the SU(2) sector of planar $\mathcal{N}=4$ SYM in the context of integrable Heisenberg spin chain  \cite{Minahan:2002ve,Beisert:2003jj,Beisert:2003yb}. 
    Followed by this, there has been number of studies put forth to establish the appearance of integrability in other lower dimensional counterparts of AdS/CFT holography, e.g., AdS$_4$/CFT$_3$ via SU(2)$\times$ SU(2) subsector of integrable SU(4) spin chain \cite{Minahan:2008hf} and AdS$_3$/CFT$_2$ via $\mathfrak{psu}(1,1|2)^2$ spin chain \cite{Borsato:2012ud, Borsato:2012ss,Sfondrini:2014via}. In the semiclassical limit, such integrable features subsequently appear in  holographically equivalent corresponding to string theories in 10-dimensional AdS bulk. However, in \cite{Kruczenski:2006pk,Arutyunov:2003uj,Arutyunov:2003za}, 
    there occurs a rather systematic and classical method to extract large category of well-known string solutions by reducing any chosen two dimensional string sigma model into one dimensional integrable Neumann-Rosochatius (NR) model. A generalised set of embedding thus deciphers various finite gap solutions of classical string-sigma model only from the solutions of integrable NR model. Being the integrable extension of Neumann model \cite{Neumann+1859+46+63}, one of the most earlier mathematical integrable models, NR system demonstrates the constrained motion of harmonic oscillator under the influence of centrifugal inverse square potential on a spherical surface of unit radius \cite{RosochatiusAA}. The Lagrangian of the original undeformed NR construction on N-dimensional unit sphere respectively assumes the form 
    \begin{equation}
        L=\frac{1}{2}\sum_{i=1}^{N}\left[x^{'2}_{i}+\frac{v_i^2}{x_i^2}-\omega^{2}_{i}x_i^2\right]-\frac{\Lambda}{2}\left(\sum_{i=1}^{N}x_{i}^{2}-1\right),
        \label{lagrangian}
    \end{equation}
with $v_i$ being a constant and $\Lambda$ is a suitable Lagrange multiplier to incorporate the spherical geometry. The corresponding set of independent constants of motion was put forward by K. Uhlenbeck in \cite{10.1007/BFb0069763} 
while studying the NR model in connection to equivariant harmonic maps into sphere. These are famously known as Uhlenbeck integrals of motion and are expressed as
\begin{equation}
    I_{i}=x_{i}^{2}+\sum_{j\neq i}\frac{1}{\omega_{i}^{2}-\omega_{j}^{2}}\left[\left(x_{i}x_{j}^{'}-x_{j}x_{i}^{'}\right)^{2}+v_{i}^{2}\frac{x_{j}^{2}}{x_{i}^{2}}+v_{j}^{2}\frac{x_{i}^{2}}{x_{j}^{2}}\right] \ .
    \end{equation}The Hamiltonian is conventionally expressed in terms of the linear combination of the integrals of motion and the conserved angular momenta conjugate to the cyclic angular coordinates, say $f_i$.
    \begin{equation}
H=\frac{1}{2}\sum_{i=1}^{2}\left(\omega_{i}^{2}I_i+\pi_{f_i}\right),
\label{hamiltonian}
\end{equation}
    The notion of such classical integrability has previously been employed as an effective way to study generic spinning and pulsating string solutions in many holographically well-explored 10-dimensional compact bulk geometries. 
    There are also a bunch of seminal studies where giant magnon \cite{Kruczenski:2006pk,Arutyunov:2003uj,Arutyunov:2003za}
     and spiky string-type solutions \cite{Ishizeki:2007we,Mosaffa:2007ty,Hayashi:2007bq} 
     are evolved on $AdS_5\times S^5$ from the solutions of such one dimensional integrable model.\\ 
     
     \noindent In this paper, we aim to implement similar NR embeddings of the rotating as well as pulsating string in I-brane background. I-brane background arised from the intersection of two orthogonal stacks of NS5 branes along $\mathbb{R}^{1,1}$ plane \cite{Itzhaki:2005tu}. From the holographic point of view, the mere significance of the stacks of parallel NS5 branes lies in the fact that the near horizon geometry of such branes accomodates the non-local Little String Theory(LST) in its worldvolume \cite{Aharony:1999ks}. 
    Moreover, the corresponding worldsheet theory captures an exactly solvable structure \cite{Callan:1992rs}. 
    This makes it plausible to reformulate classical NR integrable model from the near horizon limit of NS5 branes. 
    The intersecting configuration in our present study is acquired by stretching $k_1$ number of NS5 branes along the directions $(012345)$ and another $k_2$ number of NS5 branes along the directions $(016789)$. When all NS5 branes are coincident, the near horizon geometry reduces to
    \begin{equation*}
        R^{2,1}\times R_{\phi}\times \text{SU(2)}_{k_1}\times \text{SU(2)}_{k_2} \ ,
    \end{equation*} where $R^{2,1}$ represents a space with coordinates $x^0$, $x^1$ and a combination of two radial directions. Meanwhile $R_{\phi}$ is another combination of radial directions away from the stacks of NS5 branes and the SU(2)s correspond to the angular 3-spheres described along the directions $(2345)$ and $(6789)$. Though the relevant gauge theory descriptions are put forward in \cite{Lin:2005nh} 
    the above geometry makes the information about the system more accessible as it originates from string equations of motion. Such intersecting brane configuration preserves higher Poincare symmetry ISO(2,1) as the combination of radial directions appears in $R^{2,1}$ symmetrically with the remaining spatial direction. Plenty of research on different aspects of intersecting branes has been carried out in \cite{Hung:2006nn,Berg:2006ng,Hung:2006jh,Grisa:2006tm,Antonyan:2006pg,Antonyan:2006qy,Antonyan:2006vw,Kluson:2006wa,Kluson:2005qq,Kluson:2005eb}. 
    Furthermore, over the last two decades, people have explored a series of works on magnon-like as well as spiky solutions for rigidly rotating strings and branes in such intersecting brane backgrounds \cite{Kluson:2007st,Kluson:2008gf,Biswas:2012wu}.\\
    
    \noindent Here we illustrate an underlying integrable NR structure for the background chosen to be an intersection of equal number of orthogonal NS5 branes while supported by a fixed amount of NS-NS flux. The integrable model emerges from the $\text{SU(2)}_{k_1}\times \text{SU(2)}_{k_2}$ part when we fix the motion of the probe fundamental string along the directions of intersection. We corroborate on the families of rotating string with large angular momentum and oscillation string with small energy and large quantum oscillation number. The large charge rigid rotating strings is widely studied by enacting the integrable NR approach in order to enrich the AdS/CFT holographic realizations in various dimensions \cite{Hofman:2006xt,Kruczenski:2004wg,Minahan:2002rc,Ahn:2008hj,Hernandez:2014eta,Hernandez:2015nba,Arutyunov:2016ysi,Hernandez:2017raj,Chakraborty:2019gmt, Chakraborty:2022iuk}. 
    On the other hand, similar practice has further reproduced the oscillating strings solutions \cite{Hernandez:2018gcd,Chakraborty:2019gmt,Chakraborty:2020las} which are originally found to be more stable for higher spins than the rotating strings    \cite{Gubser:2002tv,Khan:2005fc,Khan:2003sm}. These, unlike the rotating ones, typically do not support the large $J$ limit and are conveniently expressed in terms of an adiabatic large quantum number, known as oscillation number \cite{Smedback:2004udl,Chen:2008qq,Dimov:2009rd,Engquist:2003rn,Beisert:2003xu,Beisert:2003ea}. 
    We wish to deliver some holographically well-suited rotating and pulsating string solutions by solving for the integrable model in our present context. 
    
    The rest of the paper is organised as follows. In section 2 we outline the construction of the NR like model starting with fundamental string in I-brane background. We also construct integrals of motion in this background. Section 3 is devoted to the study of rigidly rotating string with finite amount of NS-NS flux. We use a specific choice of coordinate transformation to fulfill our purpose and compute the dispersion relation between energy and angular momentum in $R_t \times S^3$. We choose the two angular momenta along $S^3$ be equal. In section 4, we compute the energy-oscillation number relation for pulsating string. In section 5 we present our conclusions.

\section{Intersecting brane background}
In this section, we revisit the near horizon geometry of the intersection of two stacks of parallel NS5 branes associated with finite amount of flux. The generic metric 
for such type IIB supergravity solution is given by 
\begin{equation}
\begin{split}
  &ds^2=-dx_0^2+dx_1^2+H_1(y)\sum_{i=2}^{5}dy_i^2+H_2(z)\sum_{j=6}^{9}dz_j^2 , \\&H_1(y)=1+\frac{k_1l_s^2}{y^2},~H_2(z)=1+\frac{k_2l_s^2}{z^2},~~~e^{2\phi}=H_1(y)H_2(z),\\&y=\sqrt{\sum_{i=2}^{5}(y_i)^2},~~z=\sqrt{\sum_{j=2}^{5}(z_j)^2},
\end{split}
\end{equation}
We now consider the near horizon limit, 
\begin{equation}
    \frac{k_1l_s^2}{y^2}\gg1,~~~ \frac{k_1l_s^2}{z^2}\gg1.
\end{equation}
Under this limit our metric takes the following form
\begin{equation}
\begin{split}
    ds^2=-dx_0^2+dx_1^2+k_1l_s^2&\left(\frac{d\rho_1^2}{\rho_1^2}+d\theta_1^2+\sin^2{\theta_1}d\phi_1^2+\cos^2{\theta_1}d\psi_1^2\right)\\&+k_2l_s^2\left(\frac{d\rho_2^2}{\rho_2^2}+d\theta_2+\sin^2{\theta_2}d\phi_2^2+\cos^2{\theta_2}d\psi_2^2\right)
    \end{split}
\end{equation}
The expressions for the flux components associated to the above metric are given by,
\begin{equation}
    B_{\phi_1\psi_1}=k_1l_s^2q\sin^2\theta_1,~~ B_{\phi_2\psi_2}=k_2l_s^2q\sin^2\theta_2,
\end{equation}
Here $q$ is the finite amount of flux introduced where for $q=1$, we get the usual I-brane background. The range of the flux parameter varies from $0<q<1$. By making the following change in variables 
\begin{equation}
k_1=k_2=N,~~r_1=\ln{\frac{\rho_1}{\sqrt{N}}},~~r_2=\ln{\frac{\rho_2}{\sqrt{N}}},~~x_0=\sqrt{N}t,~~x_1=\sqrt{N}y, 
\end{equation}
The metric and the flux components looks like \begin{equation}
\begin{split}
    ds^2=N\bigg[&-dt^2+dy^2+d\theta_1^2+\sin^2{\theta_1}d\phi_1^2+\cos^2{\theta_1}d\psi_1^2\\&\hspace{2cm}+d\theta_2^2+\sin^2{\theta_2}d\phi_2^2+\cos^2{\theta_2}d\psi_2^2\bigg]+(dr_1^2+dr_2^2),\\&B_{\phi_1\psi_1}=Nq\sin^2\theta_1,~~ B_{\phi_2\psi_2}=Nq\sin^2\theta_2.
    \end{split}
    \end{equation}
    Let us take the coordinates along the directions of intersection of the branes to be localized at $y=0$ and for$~r_1=r_2=\text{const}$, the metric reduces to
\begin{equation}
    ds^2=N\bigg[-dt^2+d\theta_1^2+\sin^2{\theta_1}d\phi_1^2+\cos^2{\theta_1}d\psi_1^2+d\theta_2^2+\sin^2{\theta_2}d\phi_2^2+\cos^2{\theta_2}d\psi_2^2\bigg]
\end{equation}
\subsection{Neumann-Rosochatius construction}
A convenient set of string embedding that can reproduce the above form of the metric may be taken as,
\begin{equation}
\begin{split}
    W_1=\sin{\theta_1}e^{i\phi_1},~~W_2=\cos{\theta_1}e^{i\psi_1},~~W_3=\sin{\theta_2}e^{i\phi_2},~~W_4=\sin{\theta_2}e^{i\psi_2},
    \end{split}
\end{equation}
At this stage we consider the generalised NR ansatz as,
\begin{equation}
	\begin{split}
	   &t=k\tau,~~~~W_1=r_1(\xi)e^{i\Theta_1(\xi)},~~~~W_2=r_2(\xi)e^{i\Theta_2(\xi)},\\&W_3=r_3(\xi)e^{i\Theta_3(\xi)},~~~~W_4=r_4(\xi)e^{i\Theta_4(\xi)},
	    \end{split}
	\end{equation}
		where $\Theta_i(\xi)=f_i(\xi)+\omega_i\tau$ and $\xi=\alpha \sigma +\beta \tau$, the radial coordinates $r_1, r_2$ and $r_3, r_4$ describing the transverse sphere parts immediately follow the constraints for spherical geometry, i.e, $r_1^2+r_2^2=1$,~~$r_3^2+r_4^2=1$.
	The Polyakov action of the string can be written as
		\begin{equation}
	   S=-\frac{T}{2}\int d\tau d\sigma \left[\sqrt{-\gamma}
\gamma^{\alpha\beta}G_{MN}\partial_\alpha X^M\partial_\beta X^N- \epsilon^{\alpha\beta}B_{MN}\partial_\alpha X^M\partial_\beta X^N\right],
	\end{equation}
where $T$ is the string tension. Substituting for the NR ansatz, the Lagrangian is subsequently expressed as,
\begin{equation}
\begin{split}
  \mathcal{L}=\frac{-TN}{2}\Bigg[k^2&+(\alpha^2-\beta^2)\sum_{i=1}^{4}\left\{ r_i'^2+r_i^2\left(f_i'^2-\frac{\beta \omega_i}{\alpha^2-\beta^2}\right)^2\right\}+\sum_{i=1}^{4}\frac{\alpha^2\omega_i^2r_i^2}{\alpha^2-\beta^2}\\&\hspace{0.3cm}+\Lambda(r_1^2+r_2^2-1)+\tilde{\Lambda}(r_3^2+r_4^2-1)+2q\alpha r_1^2\left(\omega_2 f_1'-\omega_1 f_2'\right)\\&\hspace{6.2cm}+2q\alpha r_3^2\left(\omega_4 f_3'-\omega_3 f_4'\right)\Bigg],
\end{split}
\end{equation}
where prime $(')$ denotes derivative with respect to $\xi$. Equations of motions for $f_i's$ take the following form
\begin{equation}
 f_1'=\frac{1}{\alpha^2-\beta^2}\left(\frac{A_1}{r_1^2}+\beta \omega_1-q\alpha \omega_2\right),~~
 f_2'=\frac{1}{\alpha^2-\beta^2}\left(\frac{A_2}{r_2^2}+\beta \omega_2+q\alpha \omega_1\frac{r_1^2}{r_2^2}\right)
 \end{equation}
 \begin{equation}
 f_3'=\frac{1}{\alpha^2-\beta^2}\left(\frac{A_3}{r_3^2}+\beta \omega_3-q\alpha \omega_4\right),~~
 f_4'=\frac{1}{\alpha^2-\beta^2}\left(\frac{A_4}{r_4^2}+\beta \omega_4+q\alpha \omega_3\frac{r_3^2}{r_4^2}\right)
\end{equation}
Substituting the values of $f_i'$ in the equations  of motion of radial coordinates gives,
\begin{subequations}
\begin{align}
     &(\alpha^2-\beta^2)^2 r_1''+\alpha^2(\omega_1^2+q^2\omega_2^2)r_1-\frac{A_1^2}{r_1^3}+\frac{2q\alpha A_2\omega_1r_1}{r_2^2}+\frac{2q^2\alpha^2\omega_1^2r_1^3}{r_2^2}=0\\&
    (\alpha^2-\beta^2)^2 r_2''+\alpha^2\omega_2^2r_2-\frac{A_2^2}{r_2^3}-\frac{2q\alpha A_2\omega_1r_1^2}{r_2^3}-\frac{q^2\alpha^2\omega_1^2r_1^4}{r_2^3}=0\\&
     (\alpha^2-\beta^2)^2 r_3''+\alpha^2(\omega_3^2+q^2\omega_4^2)r_3-\frac{A_3^2}{r_3^3}+\frac{2q\alpha A_4\omega_3r_3}{r_4^2}+\frac{2q^2\alpha^2\omega_3^2r_3^3}{r_4^2}=0\\&
     (\alpha^2-\beta^2)^2 r_4''+\alpha^2\omega_4^2r_4-\frac{A_4^2}{r_4^3}-\frac{2q\alpha A_4\omega_3r_3^2}{r_4^3}-\frac{q^2\alpha^2\omega_3^2r_3^4}{r_4^3}=0
\end{align}
\end{subequations}
The above equations of motions can be derived from the following lagrangian
\begin{equation}
\begin{split}
    L_{NR}=\frac{1}{2}\left[\left(\alpha^2-\beta^2\right)^2\Big\{\sum_{a=1}^{4}r_a^{'2}\Big\}-\alpha^2\Big\{\sum_{a=1}^{4}\omega_a^2r_a^2\Big\}-q^2\alpha^2\omega_2^2r_1^2-q^2\alpha^2\omega_4^2r_3^2\right. \\ \left. -\Big\{\sum_{a=1}^{4}\frac{A_a^2}{r_a^2}\Big\}-\Big\{\frac{2q\alpha\omega_1A_2r_1^2}{r_2^2}+\frac{q^2\alpha^2\omega_1^2r_1^4}{r_2^2}+\frac{2q\alpha\omega_3A_4r_3^2}{r_4^2}+\frac{q^2\alpha^2\omega_3^2r_3^4}{r_4^2}\Big\}\right].
    \end{split}
    \label{NRlagrangian}
\end{equation}
The Lagrangian evidently acquires the potentials  of the form $r^2$ and $1/r^2$ which resembles the harmonic and centrifugal barrier potential respectively. Hence, the above lagrangian can be identified as the lagrangian of the integrable one dimensional NR model with some deformations due to the presence of mixed flux in the background. We now proceed to calculate the Uhlenbeck integrals of motion for this deformed system.
\subsection{Integrals of motion}
In order to validate the construction of NR model, we need to find the integrals of motion for our deformed system and check whether they satisfy the relation $\sum_{i=1}^N\bar{I}=\alpha^2$ and $\{\bar{I}_i,\bar{I}_j\}=0~ \forall~ i,j\in \{1,2,..N\} $. Here due to the presence of flux, the expression for the undeformed Uhlenbeck integrals of motion acquire an additive term $f$ which can be calculated by solving $\bar{I}'=0$.
\begin{equation}
\begin{split}
    \bar{I}_i=\alpha^2r_i^2+\sum_{i\neq j}\frac{1}{\omega_i^2-\omega_j^2}\left[(\alpha^2-\beta^2)^2(r_ir_j'-r_jr_i')^2+\frac{A_i^2r_j^2}{r_i^2}+\frac{A_j^2 r_i^2}{r_j^2}+2f\right]
\end{split}
\label{iom}
\end{equation}
After performing the integration of $f'$ which we obtained by setting $\bar{I}'=0$, we find the expression for $f$ to be
\begin{equation}
    2f=q^2\alpha^2(\omega_2^2-\omega_1^2)r_1^2-\frac{q^2\alpha^2\omega_1^2}{r_2^2}+\frac{2 q\alpha \omega_1 A_2}{r_2^2}
    \label{f}
\end{equation}
Using (\ref{f}) in (\ref{iom}), we get the deformed independent integral of motion to be of the form,
\begin{equation}
    \begin{split}
    \bar{I}_{1}=\alpha^2(1-q^2)r_1^2+\frac{1}{\omega_1^2-\omega_2^2}\left[(\alpha^2-\beta^2)^2(r_1r_2'-r_2r_1')^2+\frac{A_1^2r_2^2}{r_1^2}+\frac{(A_2+q\alpha\omega_1)^2 r_1^2}
    {r_2^2}\right]
    \end{split}
    \label{1stintegral}
\end{equation}
Here we see that due to the presence of the flux, we get a multiplicative factor of $(1-q^2)$ with $r_1^2$ in the first term which is responsible for the modification in the relation of the sum of integrals,  $\sum_{i=1}^N\bar{I}=(1-q^2)\alpha^2$. In the absence of flux the deformed integrals of motion satisfy,$$\bar{I}_1+\bar{I}_2=\alpha^2.$$
Similarly, the Uhlenbeck integral of motion for the other sphere is given by,
\begin{equation}
    \begin{split}
    \bar{I}_3=\alpha^2(1-q^2)r_3^2+\frac{1}{\omega_3^2-\omega_4^2}\left[(\alpha^2-\beta^2)^2(r_2r_1'-r_1r_2')^2+\frac{A_3^2r_4^2}{r_3^2}+\frac{(A_4+q\alpha\omega_3)^2 r_3^2}
    {r_4^2}\right]
    \end{split}
    \label{2ndintegral}
\end{equation}
Following (\ref{hamiltonian}), the expression for Hamiltonian in terms of the deformed integrals of motion can be written as
\begin{equation}
    H=\frac{1}{2}\sum_{i=1}^{4}\left(\omega_{i}^{2}\bar{I}_i+A_i^2\right)+\frac{1}{2}\alpha^2q^2(w_2^2+w_4^2).
\end{equation}
\section{ Rotating string profile and large $J$ energy spectrum}
\subsection{Solving the model using Neumann's method}
Neumann's method is a convenient way to solve the integrable models with harmonic oscillator type potentials. It has its own significance over solving the equations of motion in the sense that it avoids the difficulty to deal with the second order differential equations. This procedure includes the reduction of the Uhlenbeck constants into the first order ordinary differential equation of an ellipsoidal coordinate $\zeta$. Such ellipsoidal coordinate satisfies the equation 
\begin{equation}
    \frac{r_1^2}{\zeta-\omega_1^2}+\frac{r_2^2}{\zeta-\omega_2^2}=0,
    \label{ellipsoidal}
\end{equation}which is associated with the parameters $\zeta_1, \zeta_2$ and $\zeta_3$. The expressions of $r_1$ and $r_2$ can be reduced from  (\ref{ellipsoidal}) as 
\begin{equation*}
    r_1^2=\frac{\zeta-\omega_1^2}{\omega_2^2-\omega_1^2},~~ r_2^2=\frac{\zeta-\omega_2^2}{\omega_1^2-\omega_2^2}.
\end{equation*}Inserting them accordingly in (\ref{1stintegral}) we get,
\begin{equation}
\begin{split}
   \frac{\zeta^{'2}}{4}=&\frac{1}{(\alpha^2-\beta^2)^2}\bigg[I_1(\omega_1^2-\omega_2^2)(\omega_1^2-\zeta)(\zeta-\omega_2^2)-\alpha^2(1-q^2)(\zeta-\omega_1^2)^2(\zeta-\omega_2^2)\\&-A_1(\zeta-\omega_2^2)^2-(A_2+q\alpha\omega_1)^2(\zeta-\omega_1^2)^2\bigg].
   \end{split}
\end{equation}
The right hand side of the above expression may be written as a third order polynomial of $\zeta$ given by
\begin{equation}
    \frac{\zeta^{'2}}{4}=-\frac{P_3(\zeta)}{(\alpha^2-\beta^2)^2}=-\frac{\alpha^2\left(1-q^2\right)}{(\alpha^2-\beta^2)^2}\left(\zeta-\zeta_1\right)\left(\zeta-\zeta_2\right)\left(\zeta-\zeta_3\right),
    \label{polynomial}
\end{equation} where the parameters $\zeta_1,\zeta_2,\zeta_3$ represents the roots of the polynomial $P_3(\zeta)$
\begin{equation*}
\begin{split}
P_3(\zeta)&=\zeta^3\left[\alpha^2(1-q^2)\right]\\&-\zeta^2\left[\alpha^2(1-q^2)(2\omega_1^2+\omega_2^2)-A_1-(A_2+q\alpha \omega_1)^2-\bar{I}_1(\omega_1^2-\omega_2^2)\right]\\&-\zeta\left[2A_1\omega_2^2-\alpha^2(1-q^2)\omega_1^2(\omega_1^2+2\omega_2^2)+2\omega_1^2(A_2+q\alpha \omega_1)^2+\bar{I}_1(\omega_1^4-\omega_2^4)\right]\\&-\left[\alpha^2(1-q^2)\omega_1^4\omega_2^2-A_1\omega_2^4-\omega_1^4(A_2+q\alpha\omega_1)^2-\bar{I}_1\omega_1^2\omega_2^2(\omega_1^2-\omega_2^2)\right]\\&\equiv\alpha^2(1-q^2)\left(\zeta-\zeta_1\right)\left(\zeta-\zeta_2\right)\left(\zeta-\zeta_3\right)
    \end{split}
    \end{equation*}
    The explicit expressions for $\zeta_1,\zeta_2$ and $\zeta_3$ can be derived by solving $P_3(\zeta)=0$. Due to excruciatingly difficult forms we do not mention them here in our work. \\
    At this stage we can impose a relevant coordinate transformation to carry out the solutions for (\ref{polynomial}). This is given as $\zeta=\zeta_{2}+(\zeta_{3}-\zeta_{2})\chi^{2}$, where $\chi$ is the new variable. Substituting $\zeta$ in terms of $\chi$ in (\ref{polynomial}), we get the following differential equation
    \begin{equation}
        \chi^{' 2}=\frac{\alpha^2 (1-q^2)}{(\alpha^2-\beta^2)^2}~(\zeta_3-\zeta_1)(1-\chi^2)(1-k+k \chi^2),
    \end{equation}
    where $k=(\zeta_3-\zeta_2)/(\zeta_3-\zeta_1)$. Solving the above differential equation, we obtain
    \begin{equation}
        \chi(\xi)=\text{cn}\left(\frac{\alpha\xi}{\alpha^2-\beta^2}\sqrt{\left(\zeta_3-\zeta_1\right)\left(1-q^2\right)}+\xi_0,k\right),
    \end{equation}
    where by rotation, we may set the integration constant $\xi_0$ to zero. It can be noted that the above integral can also be written in terms of $sn$ function with the proper choice of integration constant i.e, $\xi_0=\frac{-1}{\sqrt{1-k}}\textbf{K}\left(\frac{k}{k-1}\right)$. This immediately produces the string profile as
    \begin{subequations}
    \begin{align}
        &r_1^2(\xi)=\frac{\zeta_2-\omega_1^2}{\omega_2^2-\omega_1^2}+\frac{\zeta_3-\zeta_2}{\omega_2^2-\omega_1^2}~\text{cn}^2\left(\frac{\alpha\xi}{\alpha^2-\beta^2}\sqrt{\left(\zeta_3-\zeta_1\right)\left(1-q^2\right)},k\right),\label{r1expression}\\& r_2^2(\xi)=\frac{\zeta_2-\omega_2^2}{\omega_1^2-\omega_2^2}+\frac{\zeta_3-\zeta_2}{\omega_1^2-\omega_2^2}~\text{cn}^2\left(\frac{\alpha\xi}{\alpha^2-\beta^2}\sqrt{\left(\zeta_3-\zeta_1\right)\left(1-q^2\right)},k\right).\label{r2expression}
        \end{align}
    \end{subequations}Similar expressions for $r_3$ and $r_4$ can be found by reducing the integrals of motion (\ref{2ndintegral}) into the first order ODE of the ellipsoidal coordinate. It should be noted that in order to have consistent choice of roots, we need to take the condition, $\zeta_1 < \zeta_2< \zeta_3$, which ultimately fixes the argument of the elliptic sine function to be real with $0<k<1$. The periodicity conditions for the closed rotating string in our case read as,
\begin{equation}
    r_{1,2}(\xi+2\pi\alpha)=r_{1,2}(\xi),~~f_{1,2}(\xi+2\pi\alpha)=f_{1,2}(\xi)+2\pi m_{1,2},
\end{equation}
where $m_{1,2}$ are the integer winding numbers. These yield for $r_1^{2}$ and $r_2^{2}$ given in equations (\ref{r1expression}) and (\ref{r2expression}), following the relations
\begin{equation}
    \frac{\pi\alpha^2}{\alpha^2-\beta^2}\sqrt{(1-q^2)(\zeta_3-\zeta_1)}=n\textbf{K}\left(k\right),
    \end{equation}
    \begin{equation}
        \frac{\bf{\Pi}\left(\frac{\zeta_3-\zeta_2}{\zeta_3-\omega_1^2},k\right)\left(\omega_1^2-\omega_2^2\right)}{\left(\omega_1^2-\zeta_3\right)\textbf{K}\left(k\right)}=\frac{2\alpha}{A_1}\left[\left(1-\frac{\beta^2}{\alpha^2}\right)m_1-\frac{\beta}{\alpha}\omega_1+q\omega_2\right],
        \label{angularperiodicity1}
    \end{equation}and
      \begin{equation}
        \frac{\bf{\Pi}\left(\frac{\zeta_3-\zeta_2}{\zeta_3-\omega_2^2},k\right)\left(\omega_1^2-\omega_2^2\right)}{\left(\zeta_3-\omega_2^2\right)\textbf{K}\left(k\right)}=\frac{2\alpha}{A_2+q\alpha \omega_1}\left[\left(1-\frac{\beta^2}{\alpha^2}\right)m_2-\frac{\beta}{\alpha}\omega_2+q\omega_1\right],
         \label{angularperiodicity2}
    \end{equation}
where we have chosen $n=1$ in equations (\ref{angularperiodicity1}) and (\ref{angularperiodicity2}).
\subsection{Dispersion relation for Energy and Angular momenta}
\label{subsection 3.2}  It is obvious from the pattern of the Lagrangian in (\ref{NRlagrangian}) that the integrable model reduced from our background is nothing but a superposition of two independent NR models described under the constraints $r_1^2+r_2^2=1$ and $r_3^2+r_4^2=1$. This articulates the independent rotating dynamics of the string in each of the sphere. In course of deriving the corresponding energy-angular momenta dispersion relation, we shall switch off the dynamics of our probe string in one of the sphere in $R_t \times S^3 \times S^3$. The conserved quantities are expressed as
    \begin{equation}
       E=\frac{1}{\alpha}\int_{0}^{2\pi\alpha}d\xi\frac{\partial\mathcal{L}}{\partial(\partial_\tau t)}=2TN\pi\kappa,
       \end{equation}
       \begin{equation}
       \begin{split}
       J_1=\frac{1}{\alpha}\int_{0}^{2\pi\alpha}d\xi\frac{\partial\mathcal{L}}{\partial(\partial_\tau \Theta_1)}=\frac{E}{\kappa\left(\alpha^2-\beta^2\right)}\left[\left(A_1\beta-q^2\alpha^2\omega_1-q\alpha A_2\right)\right. \\ \left.+\frac{\alpha q}{2}\left(A_2+q\alpha\omega_1\right)\frac{\bf{\Pi}\left(\frac{\zeta_3-\zeta_2}{\zeta_3-\omega_2^2},k\right)\left(\omega_1^2-\omega_2^2\right)}{\left(\zeta_3-\omega_2^2\right)\bf{K}\left(k\right)}\right. \\ \left.+\frac{\alpha^2\omega_1\left(1-q^2\right)}{2 \left(\omega_2^2-\omega_1^2\right)}\Bigg\{2\left(\zeta_3-\omega_1^2\right)-\left(\zeta_3-\zeta_1\right)\left(2-\frac{\bf{E}\left(k\right)}{\bf{K}\left(k\right)}\right)\Bigg\}\right],
    \end{split}
    \label{J1}
    \end{equation}and
    \begin{equation}
        \begin{split}
            J_2=\frac{1}{\alpha}\int_{0}^{2\pi\alpha}d\xi\frac{\partial\mathcal{L}}{\partial(\partial_\tau\Theta_2)} =\frac{E}{\kappa\left(\alpha^2-\beta^2\right)}\left[\left(A_2\beta+\alpha^2\omega_2-\alpha A_1 q\right)\right. \\ \left.-\frac{\alpha^2\omega_2(1-q^2)}{2\left(\omega_2^2-\omega_1^2\right)}\Bigg\{2\left(\zeta_3-\omega_1^2\right)-\left(\zeta_3-\zeta_1\right)\left(2-\frac{\bf{E}\left(k\right)}{\bf{K}\left(k\right)}\right)\Bigg\}\right].
        \end{split}
        \label{J2}
    \end{equation}Doing some simple algebra from equations (\ref{J1}) and (\ref{J2}) we achieve
    \begin{equation*}
    \begin{split}
        \omega_1J_2+\omega_2J_1=\frac{E}{\kappa(\alpha^2-\beta^2)}\left[(\omega_1A_2+\omega_2A_1)\beta-\alpha q(\omega_1A_1+\omega_2A_2)+\alpha^2\omega_1\omega_2\right. \\ \left.-\alpha\beta q\omega_2^2+q(\alpha^2-\beta^2)m_2\omega_2\right]
        \end{split}
    \end{equation*}
   When $\omega_1=\omega_2=\omega$ and $m_1=m_2=m$, this eventually yields the energy in terms of angular momenta  
    \begin{equation}
        E=\kappa\omega J \left[\omega^2(1-q^2)-q m\omega \right]^{-1}=\frac{\kappa J}{\omega}\left[1+\frac{q m}{\omega}+\left(1+\frac{m^2}{\omega^2}\right)q^2+\mathcal{O}(q^3)\right],
    \end{equation} 
  
    where $J\equiv J_1+J_2$ denotes the total angular momentum. Here we have chosen some specific values of the parameters as $\alpha=1, \beta=0$, without losing the generality. Moreover, the presence of the proportionality constants $A_1$ and $A_2$ has been taken care of by using the equations (\ref{angularperiodicity1}) and (\ref{angularperiodicity2})\footnote{We use the power series expansion of the complete elliptic integral of third kind\\
   $\bf{\Pi}(n,k)=\bf{K}(k)+\frac{\pi}{4}$ $_2F_1\left[\frac{1}{2},\frac{3}{2},2,k\right]n+\frac{3\pi}{16}$ $_2F_1\left[\frac{1}{2},\frac{3}{2},3,k\right]n^2+\mathcal{O}[n^3]$ up to the leading order term.}. 
   Such solution straightforwardly yields the leading order dispersion relation for the large $J$ family of Berenstein-Maldacena-Nastase(BMN) solutions\cite{Hoare:2013ida,Hoare:2013lja}. However, the linear coefficient in the present case is evidently a power series expansion of the flux parameter $q$  along with the appearance of other parameters.

\section{NR construction for oscillating string in I-brane background}
We will venture into the analysis of NR model in our considered background for generalised oscillating string ansatz. The particular case chosen here is of the string expanding and contracting from south to north poles in one of the sphere, thereby oscillating about the centre of the sphere. The NR construction demands a suitable generalised NR ansatz which may be taken as
\begin{equation}
    t=k\tau,~~W_i=r_i(\tau)e^{i(f_i(\tau)+m_i \sigma)},~~~ i=1,2.
\end{equation}
where both the sets of radial coordinates and angular coordinates are the functions of temporal worldsheet coordinate only.
The target-space Lagrangian then assumes the expression 
\begin{equation}
    \mathcal{L}=-\frac{TN}{2}\Bigg[k^2-\dot{r_i}^2+r_i ^2(m_{i}^2-\dot{f_i}^2)+2qr_1^2(m_1\dot{f_2}-m_2\dot{f_1})\bigg].
\end{equation}The equations of motion for the angular coordinates $f_i$'s are given by
\begin{equation}
    \dot{f_1}=-qm_2+\frac{A_1}{r_1^2},~~   \dot{f_2}=-\frac{A_2}{r_2^2}+\frac{qm_1r_1^2}{r_2^2}.
    \label{fequation}
\end{equation}With the set of equations in (\ref{fequation}) in hand, the equations of motion for $r_i$ can be written as,
\begin{equation}
\begin{split}
    &\ddot{r_1}-\frac{A_1^2}{r_1^3}+(m_1^2+q^2m_2^2)r_1-\frac{2qm_1A_2r_1}{r_2^2}+\frac{2q^2m_1^2r_1^3}{r_2^2}=0,\\&\ddot{r_2}+m_2^2r_2-\frac{q^2m_1^2r_1^4}{r_2^3}-\frac{A_2^2}{r_2^3}+\frac{2qA_2m_1r_1^2}{r_2^3}=0.
    \end{split}
\end{equation}
The corresponding NR lagrangian for pulsating string is given by 
\begin{equation}
\begin{split}
    L_{NR}=\frac{1}{2}\bigg[\dot{r_a}^2-m_a^2r_a^2-&\frac{A_a^2}{r_a^2}-q^2m_2^2r_1^2-\frac{q^2m_1^2r_1^4}{r_2^2}+\frac{2qA_2m_1r_1^2}{r_2^2}\bigg].
    \end{split}
    \end{equation}
Likewise the rotating string case, here also we compute for the forms of the integrals of motion corresponding to NR structure for oscillating string. Inserting some arbitrary flux-dependent deformation $f$ in the undeformed $I$ we get,
\begin{equation}
\bar{I}_i=r_i^2+\sum_{i\neq j}\frac{1}{m_i^2-m_j^2}\left[(r_i\dot{r_j}-r_j\dot{r_i})^2+\frac{A_i^2r_j^2}{r_i^2}+\frac{A_j^2r_i^2}{r_j^2}+2f\right].
\end{equation} and eventually substituting $f$ value from the condition $\dot{I}_1=0$, we have the deformed integrals of motion
\begin{equation}
    \bar{I}_1=(1-q^2)r_1^2+\frac{1}{m_1^2-m_2^2}\left[(r_1\dot{r_2}-r_2\dot{r_1})^2+\frac{A_1^2r_2^2}{r_1^2}+\frac{(A_2-qm_1)^2r_1^2}{r_2^2}\right].
    \label{iompulsi1}
\end{equation}
The Hamiltonian corresponding to the given system can be written in terms of integrals of motion as
\begin{equation}
    H=\frac{1}{2}\left[\sum_{i=1}^{2}\left(m_i^2r_i^2+A_i^2\right)+q^2 m_2^2\right]
\end{equation}
\subsection{Solutions via Neumann method }
We employ the Neumann's method to carry out the oscillating profile of the string. Introducing ellipsoidal coordinates $\zeta$, as the root of the equation 
\begin{equation}
    \frac{r_1^2}{\zeta-m_1^2}+\frac{r_2^2}{\zeta-m_2^2}=0.
\end{equation}
   %
We derive a first order differential equation of motion of $\zeta$ from equation (\ref{iompulsi1}) as 
\begin{equation}
\begin{split}
\frac{\dot{\zeta}^2}{4}=-4 P_{3}(\zeta),
\end{split}
 \label{ellipsoideq}
\end{equation}
where $P_3(\zeta)$ stands for a third order polynomial given by
\begin{equation*}
\begin{split}
P_3(\zeta)&=\zeta^3\left[1-q^2\right]+\zeta^2\left[A_1^2+(A_2-qm_1)^2+\bar{I}_1(m_1^2-m_2^2)-(1-q^2)(m_2^2+2m_1^2)\right]\\&+\zeta \left[(1-q^2)(m_1^2+2m_2^2)m_1^2+\bar{I}_1(m_2^4-m_1^4)-2A_1^2m_2^2-2m_1^2(A_2-qm_1)^2\right]\\&+\left[\bar{I}_1m_1^2m_2^2(m_1^2-m_2^2)-(1-q^2)m_1^4m_2^2+A_1^2m_2^4+(A_2-qm_1)^2m_1^4\right].
\end{split}
\end{equation*}Thus $P_3(\zeta)$ can also be written as
\begin{equation}
  P_3(\zeta)\equiv(1-q^2)(\zeta-\zeta_1)(\zeta-\zeta_2)(\zeta-\zeta_3). 
\end{equation}
Solving the above polynomial $P_3(\zeta)=0$, explicit expressions for $\zeta_1,\zeta_2$ and $\zeta_3$ can be obtained. 
Proceeding in the similar fashion as described in the earlier section, we perform the following variable transformation $\zeta(\tau)=\zeta_{2}+(\zeta_{3}-\zeta_{2})\chi^2$, which lead us to
\begin{equation}
    \chi(\tau)=cn\left(\sqrt{(1-q^2)(\zeta_3-\zeta_1)}\tau,k\right)
\end{equation}
where, $k=(\zeta_3-\zeta_2)/(\zeta_3-\zeta_1)$. Substituting $\chi(\tau)$ in equation (\ref{ellipsoideq}), the expression of $r_1$ is found to be
\begin{equation}
    r_1^2(\tau)=\frac{\zeta_2-m_1^2}{m_2^2-m_1^2}+\frac{\zeta_3-\zeta_2}{m_2^2-m_1^2}cn^2\left(\sqrt{(1-q^2)(\zeta_3-\zeta_1)}\tau,k\right)
\end{equation}Akin to the case of rotating string, $k$ must be within the fundamental domain $0<k<1$ which implies $\zeta_1<\zeta_2<\zeta_3$. 

\subsection{Energy-oscillation number relation for equal and opposite winding}
 Following the case taken in subsection  \ref{subsection 3.2} we restrict our probe string to pulsate in one of the spherical part of the geometry $R_t \times S^3 \times S^3$ while the pulsation in another sphere is switched off. The  expressions for the conserved charges of such configuration can be given by
\begin{equation}
\begin{split}
    &E=-\int d\sigma \frac{\partial \mathcal{L}}{\partial(\partial_{\tau}t)}= (TN)\int \kappa d\sigma=2\pi TN \kappa=2\pi TN \dot{t}\\&
    J_1=2\pi TN r_1^2\left(\dot{f_1}+qm_2\right),~~~
J_2=2 \pi T N\left( r_2^2\dot{f_2}-qm_1r_1^2\right).
\end{split}
\end{equation}
We use $E=2 \pi T \mathcal{E}, ~J=2 \pi T \mathcal{J}~$ in the above equation with $J_1+J_2=J$. This gives, 
\begin{equation}
\begin{split}
    \mathcal{E}=N\dot{t},&~~~\mathcal{J}=N\left[r_1^2\left(\dot{f_1}+(m_2-m_1)\right)+r_2^2\dot{f_2}\right], 
    \end{split}
\end{equation}Here we shall adopt the conventional way of finding the energy spectrum for oscillating string in terms of oscillation number. To do this, let us utilize the first and second Virasoro constraints which can be written as
\begin{equation}
    \sum_{i=1}^2\dot{r_i}^2+r_i^2(\dot{f_i}^2+m_i^2)=\dot{t}^2=\frac{\mathcal{E}^2}{N^2},~~~~ m_1J_1+m_2J_2=0 .
    \label{virasoro}
\end{equation}When we substitute the expression for the angular coordinates $f_i$ in (\ref{virasoro}), the first Virasoro constraint takes the following form
\begin{equation}
\begin{split}
 \dot{r_1}^2+\dot{r_2}^2&+(m_1^2r_1^2+m_2^2r_2^2)+q^2\left(m_2^2r_1^2+\frac{m_1^2r_1^4}{r_2^2}\right)+\frac{1}{N^2}\left(\frac{\mathcal{J}_1^2}{r_1^2}+\frac{\mathcal{J}_2^2}{r_2^2}\right)\\&+\frac{q}{N}\left(-2\mathcal{J}_1m_2+\frac{2m_1\mathcal{J}_2r_1^2}{r_2^2}\right)=\frac{\mathcal{E}^2}{N^2}.  
\end{split}
\end{equation}
Using the geometric constraint $r_1^2+r_2^2=1$ in the above expression we get,
\begin{equation}
    \begin{split}
&\dot{r_1}^2=\frac{1}{r_1^2}\bigg[(1-q^2)(m_1^2-m_2^2)r_1^6+\bigg((m_2^2-m_1^2)+(1-q^2)m_2^2-\frac{\mathcal{E}^2}{N^2}\\&-\frac{2q}{N}(m_2\mathcal{J}_1+m_1\mathcal{J}_2)\bigg)r_1^4+\left(\frac{\mathcal{E}^2}{N^2}-m_2^2+\frac{\mathcal{J}_1^2+\mathcal{J}_2^2}{N^2}+\frac{2qm_2\mathcal{J}_1}{N}\right)r_1^2-\frac{\mathcal{J}_1^2}{N^2}\bigg].
\label{rdot}
\end{split}
\end{equation}
Here, we can see that as the value of $r_1$ varies between $0 ~\text{to}~1$, $\dot{r}_1^2$ varies between infinity to some finite value. At this stage, we shall proceed to find the oscillation number of the string which can be written as,
\begin{equation}
    \mathcal{N}_1=\int dr_1\frac{\partial \mathcal{L}}{\partial\dot{r_1}}=N\int dr_1\dot{r_1}.   
    \label{oscillation}
\end{equation}Using (\ref{rdot}), we have the oscillation number as
\begin{equation}
\begin{split}
 &\mathcal{N}_1=N\int \frac{dr_1}{r_1}
\bigg[(1-q^2)(m_1^2-m_2^2)r_1^6+\bigg((m_2^2-m_1^2)+(1-q^2)m_2^2-\frac{\mathcal{E}^2}{N^2}\\&-\frac{2q}{N}(m_2\mathcal{J}_1+m_1\mathcal{J}_2)\bigg)r_1^4+\left(\frac{\mathcal{E}^2}{N^2}-m_2^2+\frac{\mathcal{J}_1^2+\mathcal{J}_2^2}{N^2}+\frac{2qm_2\mathcal{J}_1}{N}\right)r_1^2-\frac{\mathcal{J}_1^2}{N^2}\bigg]^\frac{1}{2}.
 \end{split}
\end{equation}
In order to make it simple, we consider the case $\mathcal{J}_1=\mathcal{J}_2=\mathcal{J}$,~which in turn implies~$m_1=-m_2=m$ from the second Virasoro constraint. With this our equation reduces to
    \begin{equation}
 \mathcal{N}_1=N\int \frac{dr_1}{r_1}\sqrt{
\bigg[\left((1-q^2)m^2-\frac{\mathcal{E}^2}{N^2}\right)r_1^4+\left(\frac{\mathcal{E}^2}{N^2}-m^2+\frac{2\mathcal{J}^2}{N^2}-\frac{2qm\mathcal{J}}{N}\right)r_1^2-\frac{\mathcal{J}^2}{N^2}\bigg]}.
\end{equation}
With the proper choice of limits, the above integral can be written as 
\begin{equation} \mathcal{N}_1=N\sqrt{(1-q^2)m^2-\frac{\mathcal{E}^2}{N^2}}\int_{\sqrt{R_1}}^{1} dr_1\frac{\sqrt{(r_1^2-R_1)(r_1^2-R_2)}}{r_1}
\end{equation}
where, $R_1~ \text{and}~R_2$ are the roots of the quartic polynomial. Solving the above integral we get,
\begin{equation}
\begin{split}
   & \mathcal{N}_1=N\sqrt{(1-q^2)m^2-\frac{\mathcal{E}^2}{N^2}}\bigg[\frac{1}{2}\sqrt{(1-R_1)(1-R_2)}-\frac{1}{2}(R_1+R_2) ArcSinh{\sqrt{\frac{1-R_1}{R_1-R_2}}}\\&\hspace{7cm}-\sqrt{R_1R_2}~ArcTanh{\sqrt{\frac{R_2(R_1-1)}{R_1(R_2-1)}}}\bigg].
   \label{oscilatingnum}
\end{split}
\end{equation}
The expression for the roots $R_1 ~ \text{and}~R_2$ are as follows
\begin{equation*}
\begin{split}
  R_1=&\frac{1}{2}\bigg\{1+\frac{\sqrt{-4 \mathcal{J} m N q \left(\mathcal{E}^2+2
   \mathcal{J}^2-m^2 N^2\right)+\left(\mathcal{E}^2-m^2 N^2\right)^2+4 \mathcal{J}^4}}{\left(\mathcal{E}^2+m^2 N^2 \left(q^2-1\right)\right)}\\&\hspace{7cm}+\frac{2 \mathcal{J}^2-2 \mathcal{J} m N q-m^2 N^2 q^2}{\mathcal{E}^2+m^2 N^2 \left(q^2-1\right)}\bigg\},
   \end{split}
   \end{equation*}
   \begin{equation*}
   \begin{split}
  R_2=&\frac{1}{2}\bigg\{1-\frac{\sqrt{-4 \mathcal{J} m N q \left(\mathcal{E}^2+2
   \mathcal{J}^2-m^2 N^2\right)+\left(\mathcal{E}^2-m^2 N^2\right)^2+4 \mathcal{J}^4}}{\left(\mathcal{E}^2+m^2 N^2 \left(q^2-1\right)\right)}\\&\hspace{7cm}+\frac{2 \mathcal{J}^2-2 \mathcal{J} m N q-m^2 N^2 q^2}{\mathcal{E}^2+m^2 N^2 \left(q^2-1\right)}\bigg\}.   
\end{split}
\end{equation*}
Substituting $R_1$ and $R_2$ in (\ref{oscilatingnum}) and performing the series expansion for large $\mathcal{E}$ and small $\mathcal{J}$, our expression for oscillation number drastically simplifies to 
\begin{equation}
\begin{split}
   \mathcal{N}_1=\bigg[ \frac{1}{24q\mathcal{E}\sqrt{1-q^2}}
  &+ \frac{m^2 N^2 \sqrt{ 1 - q^2}}{12q\mathcal{E}^3}+\mathcal{O}\left(\frac{1}{\mathcal{E}^4}\right)\bigg]\bigg[- m^2 N^2 q^2 \left(3 + 4 q^2\right)\\&- 3 \mathcal{J} m N q \left(1 + 6 q^2\right)+3 \mathcal{J}^2 \left(1 + q^2\right)+\mathcal{O}(\mathcal{J}^3)\bigg] 
\end{split}
   \end{equation}
   We get the oscillation number in the series of $\mathcal{J}$ and inverse series of energy $\mathcal{E}$. Now we want to express $\mathcal{E}$ in terms of oscillation no. $\mathcal{N}_1$. Neglecting the higher order terms in $\mathcal{J}$, we get the expression for $\mathcal{E}$ as
   \begin{equation}
       \begin{split}
       & \mathcal{E} = \left\{ \frac{3 \mathcal{J}^2 \left(1 + q^2\right) - m^2 N^2 q^2 \left(3 + 4 q^2\right) - 3 \mathcal{J} m N q \left(1 + 6 q^2\right)}{
 24 q \sqrt{1-q^2} }\right\}\frac{1}{\mathcal{N}_1}\\& -\left\{\frac{48 m^2 N^2 q (1-q^2)^{\frac{3}{2}}}{
 3\mathcal{J}^2 \left(1 + q^2\right) - m^2 N^2 q^2 \left(3 + 4 q^2\right) - 3 \mathcal{J} mNq \left(1 + 6 q^2\right)}\right\}\mathcal{N}_1+\mathcal{O}(\mathcal{N}_1)^2.
       \end{split}
       \label{ENrelation}
   \end{equation}\\
   
\section{Discussions and concluding remarks}
We unravelled an exquisite appearance of the integrable Neumann-Rosochatius model for the 10d I-brane background probed by typical fundamental string. For both the general classes of rotating and oscillating string, the intersection of two orthogonal stacks of NS5 branes reduces to the integrable NR model defined on $\mathbb{R}\times S^3\times S^3$. However, our chosen ansatz has generated a superimposition of two independent NR configurations, each constrained on one of the spheres. Both of the rotating and oscillating string scenarios include relevant deformations due to the finite mixed flux supported by the background. To verify whether these deformed parts also preserve integrability of the model we  devise the Lagrangian, Uhlenbeck integrals of motion and  Hamiltonian for the system under consideration. All of these pieces, albeit contain terms representing harmonic oscillator-type and centrifugal inverse square-type potentials. Besides that, the deformed Uhlenbeck constants satisfy $\sum_{i}I_i=1$ when $\alpha=1$, for strings moving in both of the spheres independently. The Uhlenbeck constants as well as satisfy the pairwise Poisson bracket commutation relation between themselves and reproduces the corresponding Hamiltonian in the usual NR form given in (\ref{hamiltonian}). We also obtain convenient rotating and oscillating string solutions by solving the deformed NR model via standard Neumann's method. We have assumed the string rotating in one of the spheres while the rotation in another sphere remains switched off. This, for some specific parametric choices, generates the famous large $J$ BMN-like linear relation between energy and angular momentum. On the other hand, similar case for oscillating string acquires a scaling between energy and oscillation number in long string and small $J$ limit. The clear emergence of underlying classical integrability in one of the family of backgrounds with higher supersymmetry, like ours, can thus add some more strength to the study of integrability in theories more fundamental than CFT's. We wish to come back with this issue in near future.
\bibliographystyle{ieeetr}
\bibliography{ref}

\end{document}